\documentclass[prl,twocolumn,amsmath,amssymb,superscriptaddress,showpacs]{revtex4}
\newcommand{\etal}{\textit{et al}.}

\usepackage{graphicx,epstopdf}
\usepackage{mathrsfs}
\usepackage[colorlinks]{hyperref}
\usepackage{braket}
\usepackage{subfigure}

\begin{document}

%
%

\title{Nearly Optimal Measurement Schemes in a Noisy Mach-Zehnder Interferometer with Coherent and Squeezed Vacuum}

%
%

\author{Bryan T. Gard}
 \thanks{Corresponding author.} 
\email{bryantgard1@gmail.com}
\affiliation{Hearne Institute for Theoretical Physics and Department of Physics \& Astronomy, Louisiana State University, Baton Rouge, LA 70803, USA}

\author{Chenglong You}
\affiliation{Hearne Institute for Theoretical Physics and Department of Physics \& Astronomy, Louisiana State University, Baton Rouge, LA 70803, USA}

\author{Devendra K. Mishra}
\affiliation{Hearne Institute for Theoretical Physics and Department of Physics \& Astronomy, Louisiana State University, Baton Rouge, LA 70803, USA}
\affiliation{Physics Department, V. S. Mehta College of Science, Bharwari, Kaushambi-212201, U. P., India}

\author{Robinjeet Singh}
\affiliation{Hearne Institute for Theoretical Physics and Department of Physics \& Astronomy, Louisiana State University, Baton Rouge, LA 70803, USA}

\author{Hwang Lee}
\affiliation{Hearne Institute for Theoretical Physics and Department of Physics \& Astronomy, Louisiana State University, Baton Rouge, LA 70803, USA}

\author{Thomas R. Corbitt}
\affiliation{Hearne Institute for Theoretical Physics and Department of Physics \& Astronomy, Louisiana State University, Baton Rouge, LA 70803, USA}

\author{Jonathan P. Dowling}
\affiliation{Hearne Institute for Theoretical Physics and Department of Physics \& Astronomy, Louisiana State University, Baton Rouge, LA 70803, USA}

%
%

\begin{abstract}
The use of an interferometer to perform an ultra-precise parameter estimation under noisy conditions is a challenging task. Here we discuss nearly optimal measurement schemes for a well known, sensitive input state, squeezed vacuum and coherent light. We find that a single mode intensity measurement, while the simplest and able to beat the shot-noise limit, is outperformed by other measurement schemes in the low-power regime. However, at high powers, intensity measurement is only outperformed by a small factor. Specifically, we confirm, that an optimal measurement choice under lossless conditions is the parity measurement. In addition, we also discuss the performance of several other common measurement schemes when considering photon loss, detector efficiency, phase drift, and thermal photon noise. We conclude that, with noise considerations, homodyne remains near optimal in both the low and high power regimes. Surprisingly, some of the remaining investigated measurement schemes, including the previous optimal parity measurement, do not remain even near optimal when noise is introduced.
\end{abstract}
\maketitle

\section{Introduction} 

Typical parameter estimation with the use of interferometric schemes aims to estimate some unknown parameter which is probed with the input quantum states of light. In principle, the sensitivity of these measurements depends on the chosen input states of light, the interferometric scheme, the noise encountered and the detection scheme performed at the output. For a real-world example, perhaps the most sensitive of these types of interferometers are the large scale interferometers used as gravitational wave sensors \cite{Abbott2016,GEO6001,GEO6002,LIGO1,LIGO2,JapanTAMA,italyVirgo}. In general, if classical states of light are used, then the most sensitive measurement is limited to a classical bound, the shot-noise limit (SNL) \cite{lloyd2004,dd2015,dowling2002}. Despite the remarkable precision possible with classical states, improvements are still possible. Here we discuss nearly optimal measurements achievable when one considers input states of coherent and squeezed vacuum \cite{Caves1981,Smerzi07}, under many common noisy conditions and in realistic power regimes which are applicable to general interferometry. 

It is of practical interest to consider the difficulty with implementing any particular measurement scheme as every additional optical element introduces further loss into the interferometer. It has been previously shown that the parity measurement is one example of an optimal measurement for coherent and squeezed vacuum input states under lossless conditions \cite{Kaushik1}. It was also previously shown that a more involved detection scheme is optimal under photon loss \cite{onohoff09}. However, here we will discuss various common detection schemes, which are easier to implement in practice and perform nearly optimal. Discussion of a lossy MZI for Fock state inputs is also discussed in previous works \cite{focknoisemzi,focknoisemzi2}.

While there are many technical challenges in using squeezed states of light, we show here that some of the measurement techniques commonly used in a classical setup are no longer near optimal. In addition, some measurements exhibit problems with effects such as phase drift and thermal photon noise. With the goal of choosing a simple, yet well-performing measurement, we investigate homodyne \cite{Paris97}, parity measurements \cite{gerry2000,gerry2003,gerry2005,gerry2010,leeparity} and compare them to a standard intensity measurements. These measurements form a set that are either simple to implement, or are known to be optimal in the lossless case. Specifically, we confirm that, under lossless conditions, the parity measurement achieves the smallest phase variance. However, under noisy conditions, surprisingly the parity measurement suffers greatly, while the homodyne measurement continues to give a nearly optimal phase measurement. The parity measurement under losses was briefly discussed in the context of entangled coherent states by Joo \etal~\cite{Spiller11}. For the lossless case, we divide our results into two regimes, the low power regime ($|\alpha|^2<500$, e.g. small scale sensors), in which different detection schemes can lead to significantly different phase variances, and the high power regime ($|\alpha|^2>10^{5}$, e.g. large scale, devoted interferometry), where all detection schemes are nearly optimal. While our scheme may hint at applications for setups like LIGO, a much more focused analysis, outside the scope of our investigation, would be required before drawing conclusions about LIGO's performance.

\section{Method}
\begin{figure}[!htb]
\includegraphics[width=\columnwidth]{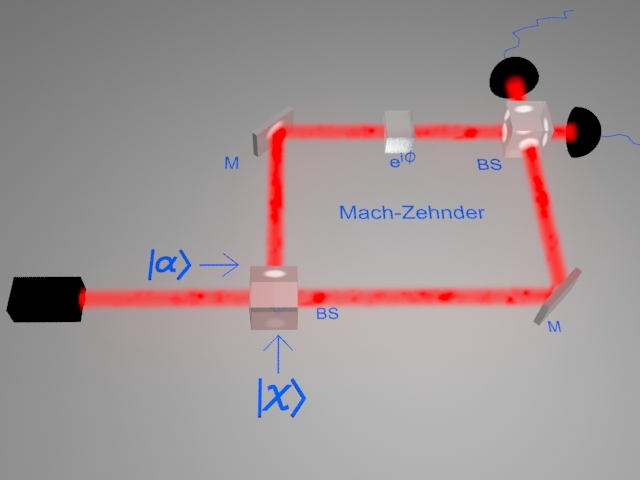}
\caption{A general Mach-Zehnder interferometer with coherent $|\alpha\rangle$ and squeezed vacuum $|\chi\rangle$ states as input. Beam splitters (BS) mix the two spatial modes, while mirrors (M) impart a phase shift, which can be safely neglected since it it common to both modes. A phase shift $\phi$ represents the phase difference between the two arms of the MZI, which can be due to a path length difference. Our goal is to estimate the unknown parameter $\phi$, which corresponds to the interaction of the quantum state with some process of interest.}
\label{fig:MZI}
\end{figure}

The interferometer considered here is a Mach-Zehnder interferometer (MZI) \cite{MZI1} as shown in Fig.~\ref{fig:MZI} and is mathematically equivalent to a Michelson interferometer. Here, an input of a coherent state $(|\alpha\rangle)$ and squeezed vacuum $(|\chi \rangle)$ is used. With this input state, it is known that the phase sensitivity can be below the SNL, typically defined as $\Delta^2 \phi_{\textrm{SNL}}=1/N$, where $N$ is the mean number of photons entering the MZI \cite{Caves1981}.

For its close connection to the parity measurement, we shall describe our states in terms of Wigner functions. One can construct any Gaussian states Wigner function directly from the first and second moments by way of,
\begin{equation}
W_\rho(\textbf{X})=\frac{1}{\pi^N}\frac{1}{\sqrt{\textrm{det}(\boldsymbol{\sigma})}}e^{-(\textbf{X}-\textbf{d})^\intercal\boldsymbol{\sigma}^{-1}(\textbf{X}-\textbf{d})}
\end{equation}
where the covariance matrix, $\boldsymbol{\sigma}=\sigma_{ij}=\langle X_i X_j + X_j X_i\rangle -2\langle X_i\rangle \langle X_j \rangle$, mean vector, $d_j=\langle X_j \rangle$ and $X_i, X_j$ are orthogonal phase space variables. 

We use this general form for our chosen input states of a coherent state $(|\alpha\rangle)$ and squeezed vacuum $(|\chi=re^{i\delta}\rangle)$ to define our states  by,
\begin{align*}
W_{\alpha}(x_1,p_1)&=\frac{1}{\pi} \exp \biggl( 2 |\alpha|  \left(\sqrt{2} (p_1 \sin\theta+x_1 \cos\theta)-|\alpha| \right)\\
& -p_1^2-x_1^2 \biggr),\\
W_{\chi}(x_2,p_2)&=\frac{1}{\pi} \exp \biggl(\sinh (2 r) \left(2 p_2 x_2 \sin \delta+\cos\delta  \left(x_2^2-p_2^2\right)\right)\\
&-\left(p_2^2+x_2^2\right) \cosh (2 r)\biggr).
\end{align*}
Here $\alpha,\theta$ are the coherent amplitude and phase, respectively while $r,\delta$ denote the squeezing parameter and phase. As the input state we consider is a product state, it can be written in terms of the product \cite{Adesso2014},
\begin{equation}
\begin{split}
W(\textbf{X})&=W_{\alpha}(x_1,p_1)\times W_{\chi}(x_2,p_2)\\
&=\frac{1}{\pi^2} e^{-p_1^2-(x_1-\sqrt{2}\alpha)^2}\times e^{-e^r p_2^2-e^{-r} x_2^2}.
\end{split}
\end{equation}
For simplicity, both states have equal initial phases, as this gives rise to the optimal phase sensitivity (discussed later) and are taken to be $\theta=\delta=0$. This simply defines the coherent state to be displaced in the $x_1$ direction and the squeezed state to be squeezed along $x_2$~\cite{bib:GerryKnight05}. The average photon number in the coherent state is $N_{\textrm{coh}}=|\alpha|^2$ and in the squeezed vacuum state $N_{\textrm{sqz}}=\sinh^2{r}$, which sets the SNL to be  $\Delta^2 \phi_{\textrm{SNL}}=1/N_{\textrm{tot}}=1/(|\alpha|^2+\sinh^2{r})$. 

The propagation of this Wigner function is accomplished by the transformation of the phase space variables through the MZI, dictated by its optical elements. These transformations are described by
\begin{eqnarray}
\textrm{BS}(1/2)=\frac{1}{\sqrt{2}}
\left(
\begin{array}{cccc}
1&0&1&0\\
0&1&0&1\\
1&0&-1&0\\
0&1&0&-1\\
\end{array} \right)
\end{eqnarray}
\begin{eqnarray}
\textrm{PS}(\phi)=
\left(
\begin{array}{cccc}
\cos(\frac{\phi}{2})&-\sin(\frac{\phi}{2})&0&0\\
\sin(\frac{\phi}{2})&\cos(\frac{\phi}{2})&0&0\\
0&0&\cos(\frac{\phi}{2})&\sin(\frac{\phi}{2})\\
0&0&-\sin(\frac{\phi}{2})&\cos(\frac{\phi}{2})\\
\end{array} \right) ,
\end{eqnarray}
where both beam splitters are fixed to be 50-50 and $\phi$ represents the unknown phase difference between the two arms of our MZI. We have chosen to use a symmetric phase model in order to simplify calculations as well as agree with previous results \cite{nori1,Hofmann}. Our goal then will be minimizing our uncertainty in the estimation of the unknown parameter $\phi$. Using these transforms, the total transform for the phase space variables is given by,
\begin{eqnarray}
\left(
\begin{array}{cc}
x_{1\textrm{f}}\\
p_{1\textrm{f}}\\
x_{2\textrm{f}}\\
p_{2\textrm{f}}
\end{array}\right)
=\textrm{BS}(1/2) \cdot \textrm{PS}(\phi) \cdot \textrm{BS}(1/2) \cdot
\left(
\begin{array}{cc}
x_{1}\\
p_{1}\\
x_{2}\\
p_{2}
\end{array} \right) ,
\end{eqnarray}
where $\{x_{j\textrm{f}},p_{j\textrm{f}}\}$ represent the phase space variables, for each mode, after propagation through the MZI.

We can also consider photon loss in the model by way of two mechanisms, photon loss to the environment inside the interferometer and photon loss at the detectors, due to inefficient detectors. Both of these can be modeled by placing a fictitious beam splitter in the interferometer with vacuum and a interferometer arm as input and tracing over one of the output modes, to mimic loss of photons to the environment \cite{photon_loss}. This linear photon loss mechanism can be modeled with the use of a relatively simple transform, since these states are all of Gaussian form. Specifically this amounts to a transform of the covariance matrix according to $\sigma_L=(1-L)\mathbb{I} \cdot \sigma +L\mathbb{I}$, $0\leq L\leq 1$ is the combined photon loss and $\mathbb{I}$ is the 4$\times$4 identity matrix. Similarly the mean vector is transformed according to $\textbf{d}_L=\sqrt{(1-L)}\mathbb{I}\cdot\textbf{d}$~\cite{onohoff09,2016APS..MAR.G1277B}.
\section{Results and Discussion}

\subsection{Quantum Cram\'{e}r-Rao Bound}
We consider an optimal measurement scheme with the meaning of saturating the quantum Cram\'{e}r-Rao bound (QCRB) \cite{cramer1946mathematical, QFI_Caves}, which gives the best phase sensitivity possible for a chosen interferometer setup and input states. This optimality is independent of measurement scheme and it remains a separate task to show which measurement scheme achieves this optimal bound \cite{Smerzi07}. In what we call the classical version of this setup, a coherent state and vacuum state are used as input. With these two input states, the best sensitivity one can achieve is bounded by the SNL, which is achievable with many different detection schemes. Many interferometer models mainly focus on analytical analysis of Fisher information~\cite{Datta11,Walmsley09} when there is loss and phase drift. While this analysis is useful in that it demonstrates a 'best case scenario', it is unknown whether the optimal detection scheme is hard to realize in an actual experimental setup. Thus, in our analysis, we are more focused on Fisher information \textit{and} how it compares with specific detection schemes, under noisy conditions.

The benefit of using squeezed vacuum in place of vacuum is then that the phase measurement can now reach below the SNL.  In order to compare various choices of measurement schemes, we not only need to calculate the various measurement outcomes, but also need to show the best sensitivity attainable with these input states. The best phase measurement one can do is given by the quantum Cram\'{e}r-Rao bound \cite{cramer1946mathematical} and is related to the quantum Fisher information (QFI, $\mathscr{F}$) \cite{QFI_Caves}, simply by $\Delta^2 \phi_{\textrm{QCRB}}=\mathscr{F}^{-1}$. For the input states of a coherent and squeezed vacuum, one can use the Schwinger representation to calculate the QFI, since these are pure states \cite{nori1,dd1a}. Another option, and the method we use here, instead utilizes the Gaussian form of the states and can be calculated directly in terms of covariance and mean \cite{QFI1,parisgaussian,gao2014}. This method applies to pure and mixed states, as long as it maintains Gaussian form. Using this formalism, the QCRB for a coherent state and squeezed vacuum into an MZI can be found to be \cite{Smerzi07, Kaushik1},
\begin{equation}
\Delta^2 \phi_{\textrm{QCRB}}=\frac{1}{|\alpha|^2 \textrm{e}^{2r}+\sinh^2(r)}.
\label{eq:qcrb}
\end{equation}
While this gives us a bound on the best sensitivity obtainable with these given input states, it does not directly consider loss or even tell us which detection scheme attains this bound. 

\subsection{Specific Measurements under lossless conditions}
Now that we have a bound on the best possible sensitivity, we now seek to show how various choices of measurement compare to this bound. We consider some standard measurement choices including single-mode intensity, intensity difference, homodyne, and parity. While each of these measurements would require a significant reconfiguration of any interferometer, it is worthwhile to show how each choice impacts the resulting phase sensitivity measurement. We utilize the bosonic creation and annihilation operators ($\hat{a}^\dagger,\hat{a}$), which obey the commutation relation, $[\hat{a},\hat{a}^\dagger]=1$. We also utilize the quadrature operators ($\hat{x},\hat{p}$) which are related to the creation and annihilation operators by the transform $\hat{a}_j=\frac{1}{\sqrt{2}}(\hat{x}_j+i \hat{p}_j)$. These quadrature operators obey a similar commutator, $[\hat{x},\hat{p}]=i$.

In terms of our output Wigner function, $\langle \hat{O}_{\textrm{sym}}\rangle=\int_{-\infty}^{\infty}O \times W(\textbf{X}) d\textbf{X}$, where ``sym" indicates that this integral calculates the symmetric ordered expectation value of the operator $\hat{O}$. Each measurement operator, $\langle \hat{O}\rangle$, gives rise to a phase uncertainty by way of $\Delta^2 \phi=\Delta^2\hat{O}/|\partial \langle \hat{O}\rangle/\partial \phi|^2$. 

Starting with the simplest measurement, an intensity measurement is given by, $\hat{O}=\langle \hat{a}^{\dagger}\hat{a} \rangle=\langle \hat{x}^2+\hat{p}^2 \rangle/2$, which is implemented by simply collecting the outgoing light, directly onto a detector. For homodyne detection, $\hat{O}=\hat{x}$ (we find the optimal homodyne measurement is taken along the $x$ quadrature). For a balanced homodyne detection scheme, one would impinge one of the outgoing light outputs onto a 50-50 beam splitter, along with a coherent state of the same frequency as the input coherent state (usually this is derived from the same source) and perform intensity difference between the two outputs of this beam splitter. While there exist other implementations of homodyne than we describe here, we choose a standard balanced homodyne scheme, for simplicity. A standard intensity difference is defined as $\hat{O}= \hat{a}^{\dagger}\hat{a}- \hat{b}^{\dagger}\hat{b}$. This particular measurement choice is also explored in Ref.~\cite{paris2015}. Parity detection is defined to be $\hat{O}=(-1)^{\langle\hat{a}^{\dagger}\hat{a}\rangle}=\pi  W(0,0)\equiv \langle \hat \Pi \rangle$. Parity detection has been implemented experimentally, though focusing on its ability for super-resolution \cite{Cohen14}. While all chosen measurements can surpass the SNL, in the lossless case, to various degrees, in order of improving phase sensitivity, single-mode intensity performs the worst, followed by intensity difference, homodyne, and finally parity. The analytical forms of each detection scheme, at their respective minima, are listed below and we confirm that, under lossless conditions, the parity measurement matches the QCRB \cite{Kaushik1}, 
\begin{equation}
\Delta^2 \phi_{\hat{\Pi}}=\frac{1}{|\alpha|^2 \textrm{e}^{2r}+\sinh^2(r)}.
\end{equation}
homodyne attains,
\begin{equation}
\Delta^2 \phi_{\hat{x}}=\frac{1}{|\alpha|^2 \textrm{e}^{2r}},
\end{equation}
 and intensity difference attains,
\begin{equation}
\Delta^2 \phi_{\hat{a}^{\dagger}\hat{a}- \hat{b}^{\dagger}\hat{b}}=\frac{\textrm{e}^{-2r}(4|\alpha|^2+(\textrm{e}^{2r}-1)^2)}{(\cosh(2r)-2|\alpha|^2-1)^2}
\end{equation}
while a single mode intensity measurement attains a minimum of,
\begin{equation}
\begin{split}
&\Delta^2 \phi_{\hat{a}^{\dagger}\hat{a}}=\\
&\frac{4|\alpha|^2\textrm{e}^{-2r}+2\cosh(2r)+4\sqrt{2} |\alpha|\sinh(2r)-2}{(\cosh(2r)-2|\alpha|^2-1)^2}.
\end{split}
\end{equation}
 We can notice that for high coherent state powers ($|\alpha|^2\gg1$), each detection scheme's leading term in its respective phase variance is given by $\Delta^2 \phi_{\textrm{all}}\approx (|\alpha|^2 \textrm{e}^{2r})^{-1}$, which is nearly optimal since the $\sinh^2(r)$ term in the QCRB is negligible compared to large $\alpha$.  From these forms then, we can say that in the low-photon-number regime ($|\alpha|^2<500$), the difference in these detection schemes can be significant, but in the high photon number regime ($|\alpha|^2>10^{5}$), there is little difference between the various detection schemes. 

\subsection{Lossy Inteferometer}
We now consider the effects of loss and calculate the lossy QCRB. This is done following the same loss procedure described previously. The lossy QCRB of this mixed state becomes \cite{onohoff09}
\begin{equation}
\begin{split}
&\Delta^2 \phi^{\textrm{Loss}}_{\textrm{QCRB}}=\\
&\frac{L(e^{2r}-1)+1}{(1-L)\lbrace|\alpha|^2e^{2r}+\sinh^2(r)[L(e^{2r}-1)+1]\rbrace}.
\label{eq:qcrbnoise}
\end{split}
\end{equation}

Note that this QCRB with loss only considers linear photon loss caused by photon loss inside the interferometer and photon loss due to inefficient detectors. In reality, there may be more specific sources of noise one needs to consider, but our method's purpose is to show a preliminary case when simple loss models are considered. We note that a measurement scheme proposed by Ono and Hofmann is exactly optimal (thus it is able to achieve the bound given by Eq.~\ref{eq:qcrbnoise}) under loss \cite{onohoff09}, but we wish to explore how simpler measurement schemes perform when compared with this bound.

\begin{figure}[!htb]
\includegraphics[width=\columnwidth]{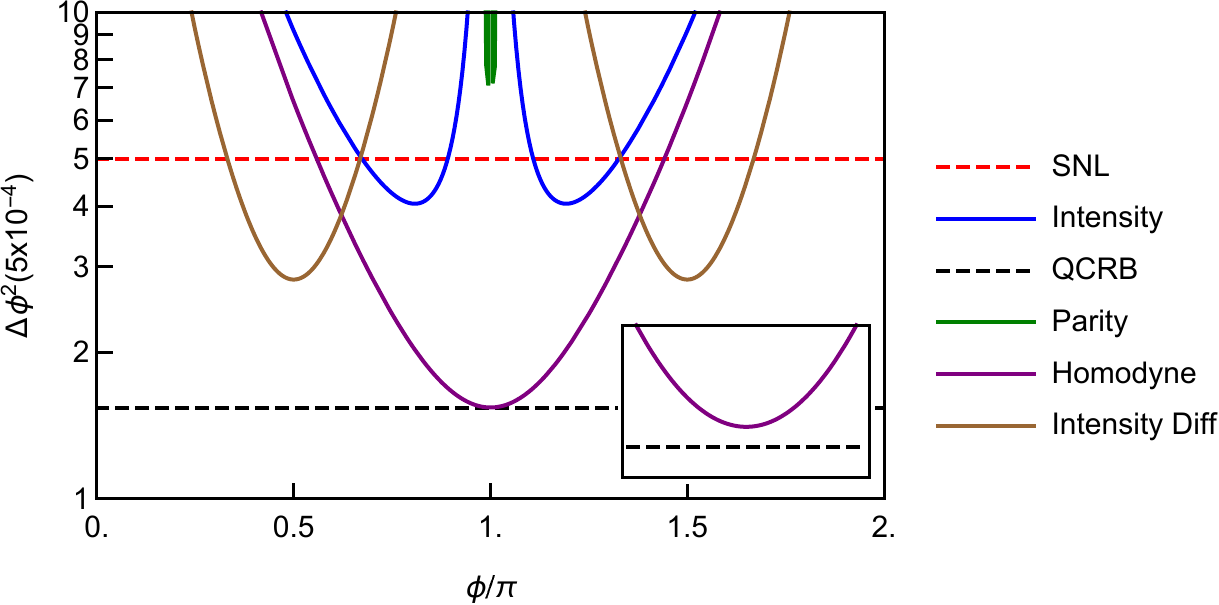}
\caption{Log plot of phase variance for various detection schemes for a coherent state and squeezed vacuum into an MZI, as a function of the unknown phase difference $\phi$. Loss parameters have been set to $L=20\%$. Input state parameters for each respective state are set to $|\alpha|^2=500$ and $r=1$. SNL and QCRB are also plotted with the same loss parameters. Note that a homodyne measurement nearly, but not exactly, reaches the QCRB as shown by the inset.}
\label{fig:phases}
\end{figure}

In the case of losses, the forms of each phase variance necessarily becomes much less appealing. For this reason, we only list the analytical form of the homodyne measurement under loss, as it is our prime candidate for a nearly optimal measurement. The phase variance of homodyne in a lossy interferometer is given by,
\begin{equation}
\Delta^2 \phi_{\hat{x}}(L)=\frac{1}{|\alpha|^2 e^{2r}}+\frac{L}{|\alpha|^2 (1-L)},
\end{equation}
where we have fixed the optimal phase to $\phi=\pi$ to obtain the phase variance minimum. We can note several interesting comparisons from this form, including the obvious $\Delta^2 \phi_{\hat{x}}(L) \geq \Delta^2 \phi_{\hat{x}}$ and $\Delta^2 \phi_{\hat{x}}(0)=\Delta^2 \phi_{\hat{x}}$. However, if we investigate Eq.~\ref{eq:qcrbnoise} for high powers (large $|\alpha|$), we find,
\begin{equation}
\begin{split}
\Delta^2 \phi^{\textrm{Loss}}_{\textrm{QCRB}}&=\frac{1}{|\alpha|^2 e^{2r}}+\frac{L}{|\alpha|^2 (1-L)}+\mathcal{O}\left(\frac{1}{|\alpha|^4}\right)\\ \nonumber
&\approx \Delta^2 \phi_{\hat{x}}(L) .  \nonumber
\end{split}
\end{equation}
This expansion illustrates the fact that homodyne is nearly optimal and approaches the QCRB in the large power limit. One can see in Fig.~\ref{fig:phases}, which shows when loss is considered, parity detection suffers greatly, while other detection schemes are still able to achieve sub-SNL phase variances. In all but the intensity and parity measurement schemes, the optimal phase (the point at which each curve achieves its minimum) has a constant value and therefore should not prove overly difficult to stabilize. In the case of intensity and parity measurement however, this optimal phase depends on both the squeezing strength $r$ and the amplitude of the coherent state $|\alpha|$. Therefore, fluctuations in the source will actually affect the optimal phase setting and in general degrade the phase measurement in this measurement scheme. Note that, in practice, typical experiments use an offset to remain near these optimum values, but purposely remain slightly away from the minimum, due to noise considerations. At this point we can note, that current technological limits enforce $|\alpha|^2\gg \sinh^2(r)$, as generally it is relatively easier to increase laser power, than to increase squeezing power. Just as it was in the lossless case, under lossy conditions then,  Fig.~\ref{fig:phases} shows that homodyne remains nearly optimal in the low power regime. In contrast, the previously optimal measurement, parity, is now not able to even reach sub-SNL.

\begin{figure}[!htb]
\includegraphics[width=\columnwidth]{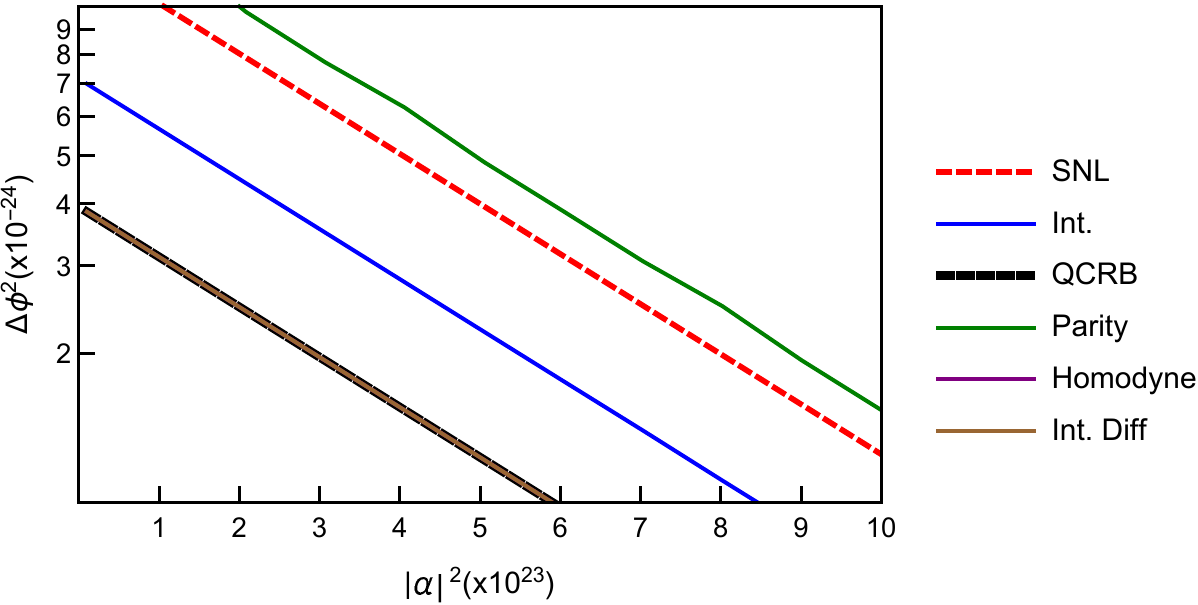}
\caption{Log plot of phase variance for various detection schemes for a coherent state and squeezed vacuum into an MZI, as a function of the average coherent photon number, $|\alpha|^2$, shown in the very large power regime. Total loss parameters have been set to $L=20\%$. We have assumed one can set the control phase to its optimal value, to obtain the best phase variance in each measurement choice. Squeezing strength in the squeezed state is set to $r=1$. Note that Parity is now not able to achieve even sub-SNL, due to loss, while homodyne and intensity difference quickly approach the QCRB (appear on top of one another). SNL and QCRB are also plotted with the same lossy parameters.}
\label{fig:nums}
\end{figure}

We can also plot the phase variance as a function of average photon number, shown in Fig.~\ref{fig:nums} for large powers, which can be related to the light's optical frequency and power by $|\alpha|^2=P/(\hbar \omega_0)$ \cite{GEO_bound}. The phase variances shown in Fig.~\ref{fig:nums} are at their respective minima in terms of optimal phase.  In this form, it's clear that a parity measurement suffers greatly, under lossy conditions and at high powers. Parity may also be difficult to implement in certain inteferometric setups as it either involves number counting (which is not feasible at very large powers) or several homodyne measurements \cite{plick2010}. Alternatively, a single homodyne measurement is nearly optimal in this lossy case, still only requires measurement on a single mode, is simpler to implement than parity, and is not nearly as sensitive to loss. We note that while homodyne appears to meet the QCRB in Fig.~\ref{fig:phases}, it actually doesn't exactly reach the QCRB (as indicated in the inset of Fig~\ref{fig:phases}). While intensity difference is also close in phase variance to a homodyne measurement (when $|\alpha|^2>100$) it requires utilization of both output modes for phase measurement, which may not be feasible in some setups. We note that while we have chosen typical parameters for $|\alpha|$ and $r$, the trend of homodyne achieving near optimal measurement generalizes to other parameter choices as well.

\subsection{Phase Drift}
Returning to Fig. \ref{fig:phases}, it is clear at which value of phase the various measurements attain their lowest value. It is this value of phase that one attempts to always take measurements at with the use of a control phase inside the interferometer. The width of each of curve then can be interpreted as the chosen measurement scheme's resistance to phase drift. The mechanism of phase drift comes about due to the limited ability to set control phases in the interferometer with infinite precision. In general, the control phase value will vary around the optimal phase setting. For this reason we aim to show this phase drift in a more rigorous way. We therefore will use the analytical forms of the various measurement phase variances, as a function of unknown phase $\phi$. We simulate phase drift by computing a running average of the phase variance, with a pseudo-randomly chosen phase, near the optimal phase, for each measurement. This pseudo-random choice is made from a Gaussian distribution, whose mean is fixed at each measurements respective optimal phase choice and has a chosen variance of $\sigma=0.15$. As predicted in the previous discussion, this gives a clearer picture of each measurement's behavior under phase drift. For simplicity, we focus on the lossless case for this treatment of phase drift.

\begin{figure}[!htb]
\centering
\begin{subfigure}
{\includegraphics[width=0.8\columnwidth]{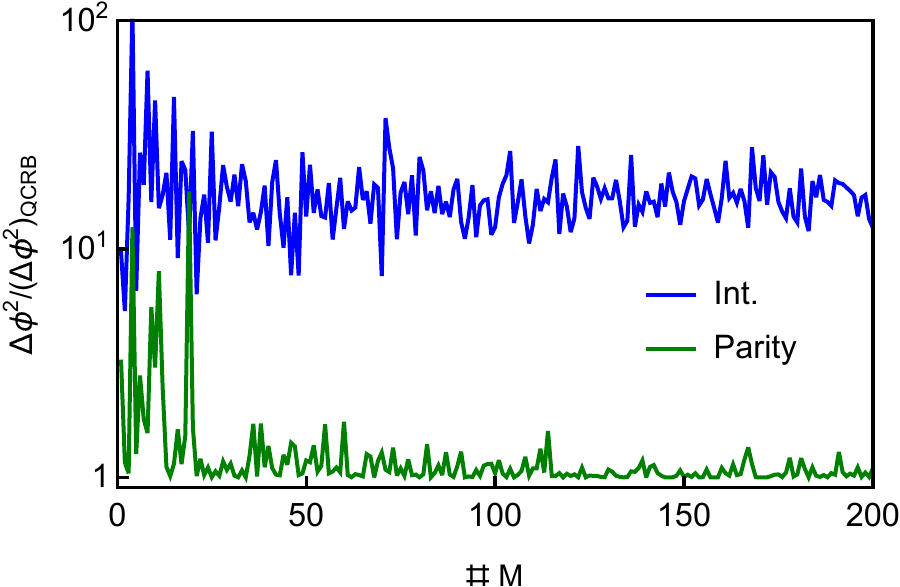}}
\end{subfigure}
~
\begin{subfigure}
{\includegraphics[width=0.8\columnwidth]{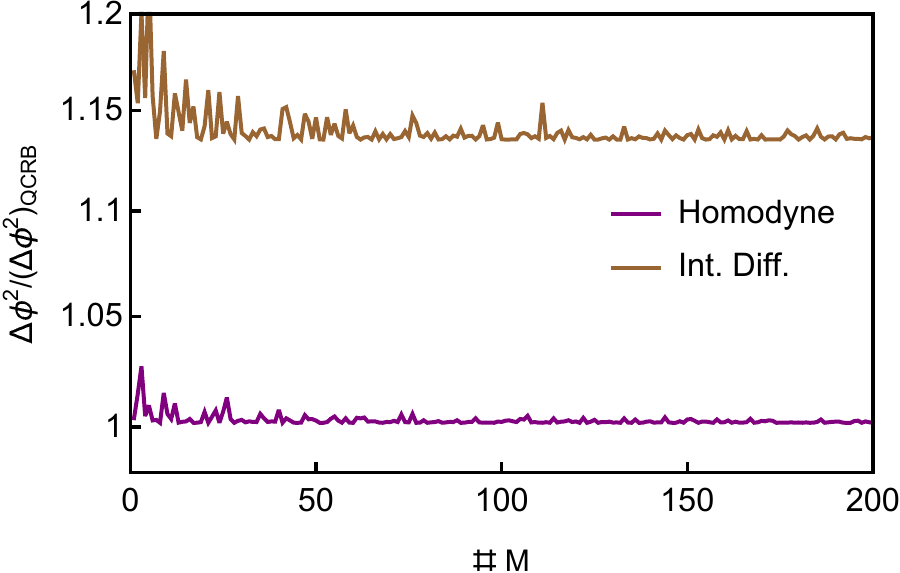}}
\end{subfigure}
\caption{Log plots of phase variance ratio to the lossless QCRB as a function of number of measurements ($M$) under phase drift noise only. For all plots shown, $|\alpha|^2=100$, $r=1$ and the standard deviation of the chosen Gaussian distribution is fixed to $\sigma=0.15$. Note that parity and intensity measurement remain noisy even after ~200 averaged measurements, where homodyne and intensity difference approach their optimal phase variance quickly in $M$.}
\label{fig:phasedrift}
\end{figure}

Shown in Fig.~\ref{fig:phasedrift}, we see the phase variance ratio to the QCRB for each measurement scheme, as a function of the number of measurements. As the number of measurements is increased, the phase variance asymptotes to the ideal measurement case, given by the phase variance at the optimal phase. This is an illustration of the law of large numbers, as the number of measurement increases, each scheme approaches its true average. However, as is clear from Fig.~\ref{fig:phasedrift}, each scheme approaches its average at significantly different rates. We can see that parity performs fairly poorly, as compared to the other measurement schemes. In the case of intensity, homodyne, and intensity difference measurements, it's clear that homodyne and intensity difference attain a small phase variance, while also being more tolerant of phase drift. This confirms a special case of Genoni \etal~ \cite{Paris11}, who showed that homodyne measurement is resistant to phase diffusion in pure Gaussian states. In principle, all of the different measurement schemes will each attain their respective phase variance minimum, as the number of measurements increases to infinity, but it is instructive to see how quickly a finite number of measurements approaches the ideal phase variance minimum.

\subsection{Thermal Photon Noise}
\begin{figure}[!htb]
\includegraphics[width=0.9\columnwidth]{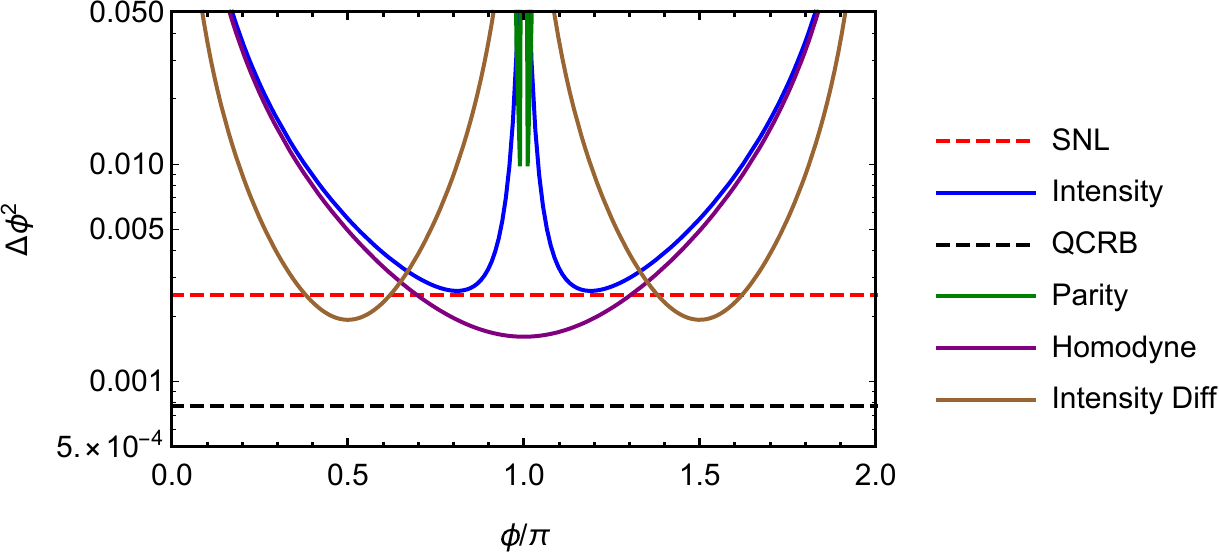}
\caption{Log plot of phase variance of the various detection schemes, with introduction of thermal photon noise into both interferometer arms, of total average photon number of $n_{\textrm{th}}=1$. Strength of the two input sources are set to $|\alpha|^2=500$, $r=1$. Note that homodyne loses most of its previous advantage over intensity difference but remains the superior measurement choice.}
\label{fig:thermal}
\end{figure}

In addition to photon loss, detector efficiency, and phase drift, we also model the inevitable interaction with thermal photon noise from the environment. This is accomplished much in the same way as a photon loss model, but here we consider a thermal photon state incident on a fictitious beam splitter, on both arms of the interferometer, and trace out one of its output modes. This allows a tunable amount of thermal photon noise (by changing the average photon number in the thermal state), into the interferometer. The effects of this unwanted thermal photon noise, to the various measurement's phase variance, is shown in Fig.~\ref{fig:thermal}. From this, we can see that even in the regime of a relatively low photon number of thermal photon noise, such noise significantly degrades the phase variance of each scheme, but drastically affects the parity scheme, making it significantly above the SNL. The SNL and QCRB under this noise model do not directly incorporate the additional thermal photons. Also in this regime, a standard single-mode intensity measurement now does not acheive sub-SNL phase variance, but homodyne and intensity difference continue to reach sub-SNL. We also note that the advantage of homodyne over intensity difference measurement is significantly decreased in the presence of thermal photon noise, but homodyne still maintains its superiority. The introduction of larger average thermal photon number continues to degrade all measurements so that they no longer beat the SNL, but this example showcases their behavior under this noise model. It should be noted that in the optical regime, the occupation of a thermal state, at room temperature is approximately $n_{\textrm{th}}\approx10^{-20}$ and therefore, some interferometric schemes do not deal with significant contribution from this model of thermal photon noise, but experiments in the microwave frequencies can have $n_{\textrm{th}}\approx1$, where this model is more applicable.

We recommend that a homodyne measurement is the simplest, nearly optimal measurement choice for a setup as discussed here. Homodyne is a typical measurement choice in interferometer experiments, as well as being a single mode measurement, likely resistant to photon loss, detector efficiency, and phase drift. It shows its main benefits in the low power regime, but performs nearly optimal in both the low and high power regimes. 

\section{Conclusion}

In this paper, we have seen the performance of many common interferometric measurement schemes. While all are able to achieve a sub-SNL phase variance measurement in the lossless case, for the choice of a coherent and squeezed vacuum input state, all are outperformed by a homodyne measurement when loss is introduced. While these measurements each come with their own challenges in implementing, we have shown that each measurement's performance can vary significantly under different noise models. We have also shown that in the high-photon regime, with loss, most measurement schemes approach the QCRB except for parity which suffers significantly. Our results may imply that simpler measurement schemes are overall appealing when using large powers. The behavior of each measurement scheme under phase drift and thermal photon noise is also discussed, and we find that homodyne and intensity difference measurement behave best within these models. This should be expected as both homodyne and intensity difference measurements operate in a similar way, subtracting intensities between two modes, removing common noise sources.  Therefore, when considering ease of implementation as well as near optimal detection, we conclude that homodyne is nearly optimal under loss, phase drift, and thermal photon noise, for the specific choice of input states of coherent and squeezed vacuum and in both power regimes.

%
%
\begin{acknowledgements}
B.T.G. would like to acknowledge support from the National Physical Science Consortium \& National Institute of Standards and Technology graduate fellowship program as well as helpful discussions with Dr. Emanuel Knill at NIST-Boulder. C.Y. would like to acknowledge support from an Economic Development Assistantship from the National Science Foundation and the Louisiana State University System Board of Regents. D.K.M. would like to acknowledge support from University Grants Commission, New Delhi, India for Raman Fellowship. T.R.C. would like to acknowledge support from the National Science Foundation grants PHY-1150531. This document has been assigned the LIGO document number LIGO-P1600084. J.P.D would like to acknowledge support from the Air Force Office of Scientific Research, the Army Research Office, the Boeing Corporation, the National Science Foundation, and the Northrop Grumman Corporation. We would all also like to thank Dr. Haixing Miao and the MQM LIGO group for helpful discussions. All authors contributed equally to this work.
\end{acknowledgements}

\appendix
\section{Declarations}
\subsection{Appendix 1: Measurement}
We have shown that different choices of measurements lead to varied ability to perform parameter estimation. Here we show the details of each detections analytical calculation for the ideal, lossless case. For each detection choice, $\hat{O}$, we need to calculate 
\begin{equation}
\begin{split}
\Delta^2 \phi&=\Delta^2\hat{O}/|\partial \langle \hat{O}\rangle/\partial \phi|^2\\
&=(\langle \hat{O}^2\rangle-\langle \hat{O}\rangle^2)/|\partial \langle \hat{O}\rangle/\partial \phi|^2
\end{split}
\end{equation}
that is, we need the variance of the chosen measurement and the derivative of its first moment. Therefore, in general we need the first and second moments for each chosen detection scheme. We also need the Wigner function at the output, this is obtained by following the transformation of phase space variables described in the text and results in a final output Wigner function of,
\begin{equation}
\begin{split}
&W(x_1,p_1,x_2,p_2)=\\
&\frac{1}{\pi^2}\textrm{Exp}\{-\frac{1}{2}e^{-2r}[p_2^2+x_1^2+e^{4r}(p_1^2+x_2^2)\\
&+e^{2r}(p_1^2+p_2^2+x_1^2+x_2^2+4|\alpha|^2\\
&+4\sqrt{2}|\alpha|(x_2\cos{(\phi/2)}-p_1\sin{(\phi/2)}))\\
&+2e^{2r}\sinh{r}((p_2^2-x_1^2+e^{2r}\cos{\phi}(p_1-x_2)(p_1+x_2))\\
&+2\sin{\phi}(p_2x_1+e^{2r}p_1x_2))]\},
\end{split}
\label{eq:wigout}
\end{equation}
where subscripts label spatial modes, $\alpha$ is the amplitude in the coherent state and $r$ the squeezing strength in the squeezed state and we have chosen $\theta_{\textrm{coh}}=\delta_{\textrm{sqz}}=0$ for simplicity.
For an intensity measurement, in terms of phase space quadrature operators $\hat{x},\hat{p}$, for the second moment, we have,
\begin{equation}
\begin{split}
\langle (\hat{a}^\dagger\hat{a})^2\rangle&=\langle (\hat{a}^\dagger\hat{a})^2\rangle_{\textrm{sym}}-\langle \hat{a}^\dagger\hat{a}\rangle-\frac{1}{2}\\
&=\int_{-\infty}^{\infty}\left[\frac{1}{4}(x^2+p^2)^2 -\frac{1}{2}(x^2+p^2)\right]W(\textbf{X}) d\textbf{X}-\frac{1}{2}
\end{split}
\end{equation}
where ``sym" denotes the symmetric operator form which is calculated from $\langle(\hat{a}^\dagger\hat{a})^2\rangle_{\textrm{sym}}=\int_{-\infty}^{\infty}\frac{1}{4}(x^2+p^2)^2 \times W(\textbf{X}) d\textbf{X}$ and $W(\textbf{X})$ is the output Wigner function, given by Eq.~\ref{eq:wigout}. For the variance calculation then,
\begin{equation}
\begin{split}
\Delta^2(\hat{a}^\dagger\hat{a})&=\langle (\hat{a}^\dagger\hat{a})^2\rangle-\langle \hat{a}^\dagger\hat{a}\rangle^2\\
&=\int_{-\infty}^{\infty}\left[\frac{1}{4}(x^2+p^2)^2 -\frac{1}{2}(x^2+p^2)\right]W(\textbf{X}) d\textbf{X}-1/2\\
&-\left[\int_{-\infty}^{\infty}\frac{1}{2}(x^2+p^2)W(\textbf{X}) d\textbf{X}-1/2\right]^2
\end{split}
\end{equation}
For homodyne detection, since the optimal homodyne measurement is along $\hat{x}$, we simply have,
\begin{equation}
\Delta^2\hat{x}=\int_{-\infty}^{\infty}x^2 W(\textbf{X})d\textbf{X} -\left(\int_{-\infty}^{\infty}x W(\textbf{X})d\textbf{X}\right)^2
\end{equation}
For intensity difference we have,
\begin{equation}
\begin{split}
&\Delta^2[(\hat{a}^\dagger\hat{a})_1-(\hat{a}^\dagger\hat{a})_2]=\\
&\int_{-\infty}^{\infty}\frac{1}{4}\left[(x_1^2+p_1^2)-(x_2^2+p_2^2)\right]^2W(\textbf{X}) d\textbf{X}\\
&-\left\{\int_{-\infty}^{\infty}\frac{1}{2}\left[(x_1^2+p_1^2)-(x_2^2+p_2^2)\right]W(\textbf{X}) d\textbf{X}\right\}^2
\end{split}
\end{equation}
and finally for parity measurement,
\begin{equation}
\Delta^2\hat{\Pi}=1-[\pi~W(0,0)]^2
\end{equation}
utilizing $\langle \hat{\Pi}^2\rangle=1$ and $W(0,0)$ is the value of the Wigner function at the origin in phase space.

\subsection{Competing Interests}
All authors declare that they have no competing interests.

\subsection{Authors’ contributions}
B.T.G. drafted the manuscript and wrote code for simulations. C.Y. and D.K.M. assisted with simulations and interpretation of results. R.S., H.L., T.R.C., and J.P.D. edited manuscript, guided simulation goals and shaped project purpose. All authors read and approved the final manuscript.
%
%
\bibliography{bib}
\end{document}